\documentclass[conference, compsocconf]{IEEEtran}

\ifCLASSINFOpdf

\else

\fi

\usepackage{graphicx}
\usepackage{cite}
\usepackage{amsmath,amssymb,amsfonts}
\usepackage{graphicx}
\usepackage[caption=false]{subfig}
\usepackage{textcomp}
\usepackage{xcolor}
\usepackage{gensymb} % used for writing \degree
\usepackage{algorithm}
\usepackage{algorithmic}
\usepackage[multiple]{footmisc}
\let\labelindent\relax

\usepackage[shortlabels]{enumitem}
\usepackage[font=small,skip=0pt]{caption}
\usepackage{comment} % used for comment
\usepackage{float} % used for subfloat
\usepackage{longtable} % used for longtable
\usepackage{enumitem} % used for removing indentation for bullet points

\begin{document}
%\title{Fine-grained Conflict Detection of IoT Services}
\title{Impact Conflict Detection of IoT Services in Multi-resident Smart Homes} %\vspace{-16mm}}


\author{\IEEEauthorblockN{Dipankar Chaki}
\IEEEauthorblockA{School of Computer Science\\
The University of Sydney\\
Sydney, Australia\\
dipankar.chaki@sydney.edu.au}
\and
\IEEEauthorblockN{Athman Bouguettaya}
\IEEEauthorblockA{School of Computer Science\\
The University of Sydney\\
Sydney, Australia\\
athman.bouguettaya@sydney.edu.au}
\and
\IEEEauthorblockN{Abdallah Lakhdari}
\IEEEauthorblockA{School of Computer Science\\
The University of Sydney\\
Sydney, Australia\\
abdallah.lakhdari@sydney.edu.au}
}

\maketitle

\begin{abstract}
We propose a novel impact conflict detection framework for IoT services in multi-resident smart homes. The proposed impact assessment model is developed based on the integral of a signal deviation strategy. We mine the residents' previous service usage records to design a robust preference estimation model. We design an impact conflict detection approach using temporal proximity and preferential proximity techniques. Experimental results on real-world datasets demonstrate the effectiveness of the proposed approach.
\end{abstract}

\begin{IEEEkeywords}
IoT services; Multi-resident smart homes; Service impact assessment; Signal deviation strategy; proximity technique; Conflict detection
\end{IEEEkeywords}

\IEEEpeerreviewmaketitle

\section{Introduction}
Internet of Things (IoT) is the umbrella term covering everyday objects (a.k.a. things) connected to the Internet. These are usually equipped with ubiquitous intelligence \cite{nauman2020multimedia}. The rapid advancement of the underlying technologies such as wired sensor networks, wireless sensor networks, and Radio Frequency Identification (RFID) tags provide these ``things" with augmented capabilities such as networking, actuating, and sensing \cite{marwedel2021embedded}. IoT technologies enable many innovative and pioneering applications, such as smart campuses, smart offices, smart cities, intelligent transport systems, and smart grids. \emph{Smart home} is another cutting-edge application domain of IoT. Any regular home fitted with IoT devices is defined as a smart home. These IoT devices are attached to everyday ``things" to monitor usage patterns. For example, a sensor (i.e., an IoT device) attached to a cup may monitor a resident's tea-cup usage patterns. The purpose of a smart home is to provide its residents with \textit{convenience} and \textit{efficiency} \cite{chaki2021adaptive}.

The concept of IoT is congruent with the \emph{service paradigm} \cite{bouguettaya2021internet}. Each ``thing" has a set of \emph{functional} and \emph{non-functional} (a.k.a. quality of service) properties. In this regard, we leverage the service paradigm as a framework to define the functional and non-functional properties of smart home devices as \emph{IoT services} \cite{huang2016service}. For instance, an Air-conditioning unit (AC) in a smart home is represented as an AC service. The functional property of the AC service is to control the temperature inside an ambient environment. Examples of non-functional properties include AC types (i.e., split-system, ducted, portable), capacity (i.e., a 2.6kW unit may cool the room of size $10m^2$ to $20m^2$), fan speed, fan direction, noise level, sleep mode, and dehumidifier mode.

We identify two categories of smart homes: (i) \emph{Single-resident} and (ii) \emph{Multi-resident}. The distinction between these two types is important since the nature of service conflicts would differ for each type. This paper focuses on IoT service conflicts that occur in multi-resident smart homes. Residents may have different preferences for using a service, which may cause \textit{IoT service conflict} \cite{chaki2020conflict}. For example, a resident may prefer 20\degree C AC temperature in the living room. At the same location and time, another resident may have a different AC temperature preference, such as 25\degree C. Hence, a service conflict may occur as the AC cannot satisfy the preferences of both residents simultaneously. In this context, conflicts on a \emph{single} IoT service are intuitive. Furthermore, conflicts may occur on \emph{multiple} IoT services when residents have different preferences for using them. For example, a resident may prefer watching a movie in a theater-like atmosphere. When the resident turns on the TV, they dim the light, close the curtain, and put the AC at 20\textdegree C. At the same time, another resident enters the same room and opens the window blind service since they prefer sunlight during the day. Opening the blind may ruin the first resident's movie-watching experience as it increases the overall illumination of the ambient environment. Although the functionalities of window blind and light services are not directly dependent, their operations may indirectly interfere via an environment property (i.e., illumination). We define it as \emph{service impact conflict} since the simultaneous enactment of two or more IoT services may affect the experience of the service users in a multi-occupant smart home. Such conflicts may create a less comfortable home environment if not handled properly. This may hamper the ultimate goal of smart homes, which is to make occupants' lives more convenient, efficient, secure, and comfortable. Indeed, a prerequisite to providing convenience that suits all occupants is to \emph{detect} conflicts first and then \emph{resolve} them. Therefore, this paper's main focus is on \emph{detecting service impact conflicts}.

The need to provide convenience in smart homes has, of course, not gone unnoticed. A few works focus on designing comfortable smart homes without considering IoT service conflicts \cite{huang2018convenience, hua2019riot, huang2018discovering}. Existing literature on conflict management is scarce in ambient intelligence systems. To date, most of the conflict detection frameworks consider multi-user conflicts regarding an individual IoT service \cite{chaki2020conflict, hua2022copi, chaki2020fine}. Compared to the conflict detection of an individual IoT service, it is generally more difficult to detect the impact conflict of multiple IoT services. Only a limited amount of existing literature focuses on impact conflict detection considering environmental properties. An object-oriented framework is proposed in \cite{nakamura2013considering} to express the interaction of environment properties and services. Matsuo et al. used the notion of effect direction to extend the previous object-oriented framework \cite{matsuo2008verifying}. Resource-locking techniques are introduced to formalize environment interactions in \cite{kolberg2003compatibility, du2008considering}, and the dependency between the control and the environment is modeled with a goal-oriented framework. However, we identify three research gaps that this work fills:

\begin{itemize}[leftmargin=*]
    \item \textbf{Impact Quantification:} Existing frameworks do not consider the degree (or amount) of environmental impact. They only deal with the dependency between an operation (i.e., opening a window blind) and the environment (i.e., illumination) but do not quantify the impact posed by the dependency. For example, they can only identify the relationship between blind service and illumination but cannot calculate the effect of opening the blind on ambient illumination. Thus, we can not distinguish whether the impact is small or big.
    
    \item \textbf{Preference Estimation:} The notion of conflict depends on the occupants' preferences. In practice, some residents are very flexible, and some are very strict in terms of preferences. For instance, one resident may not tolerate any external sound while studying (i.e., preferred sound range 0-20 decibels), whereas another resident does not bother about any sound while studying. Turning on the TV service impacts the ambient sound; it may create a conflict for the first resident but may not create a conflict for the second resident. Existing approaches do not consider preferences while identifying conflicts.
    
    \item \textbf{Determinism:} Conflicts are handled deterministically by current frameworks; they either happen or not. However, impact conflicts are not always certain to occur because of the dynamics of user behavior and context (e.g., location, time, or weather). Therefore, detecting impact conflicts and their likelihood of happening not only provides additional information for conflict resolution but also accurately reflects the severity of the conflicts.
    %Although the advantage of probabilistic approach is acknowledged in identifying conflicts on an individual service, no published work that explicitly outputs impact conflict situations with related probabilities.
\end{itemize}

We propose a novel approach for conflict detection of IoT services considering impact and residents' preferences. The impact is quantified from residents' current service requirements, and preferences are estimated from historical interactions. The proposed impact assessment model is developed based on the integral of signal deviations strategy adopted from Signal Temporal Logic (STL) \cite{donze2010robust}. We then employ a distance-based clustering algorithm, DBSCAN, to extract the residents' service usage preferences. Finally, we design the impact conflict detection approach considering preference scores and impact scores. The proposed approach employs preferential proximity and temporal proximity techniques to calculate the likelihood of conflicts. The contribution of this paper is threefold:

\begin{itemize}[nosep]
    \item An impact assessment model of IoT services adopting the concept of signal deviation integration from Signal Temporal Logic (STL). 
    \item A preference estimation model based on distance-based clustering.
    \item An impact conflict detection approach that employs preferential proximity and temporal proximity strategy to compute conflict likelihood.
\end{itemize}

The rest of the paper is structured as follows. Section II illustrates the motivation scenario to explain the challenges for service impact conflict detection. In section III, we introduce a set of terminologies and concepts that are used to formulate the problem. Section IV describes the proposed conflict detection framework. Section V presents our experiments to evaluate the proposed approach. Section VI summarizes the related work on impact conflict detection. Section VII discusses the constraints of our framework and concludes the paper with future work.

\section{Motivation Scenario}
%\vspace{-1mm}
We discuss the following two scenarios to illustrate the notion of service impact conflict in a multi-tenant environment. In this work, we consider four environmental properties: (i) temperature, (ii) illumination, (iii) sound, and (iv) humidity.\looseness=-1

\begin{figure*}[t] 
    \centering
    \includegraphics[width=\textwidth]{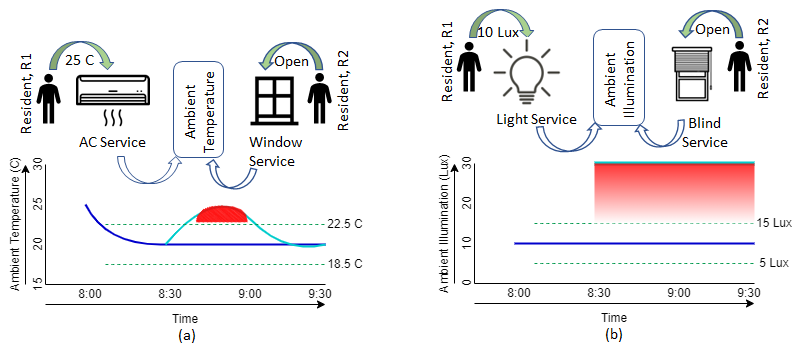}
    % \vspace{-4mm}
    \caption{Service impact conflict: (a) progressive change, (b) instantaneous change.}
    \label{motivation}
    % \vspace{-4mm}
\end{figure*}

\textit{Scenario 1:}
Suppose two services (AC service, window service) are located in the living room (Fig. \ref{motivation}(a)). Assume that the current room temperature is 25\degree C. At 8:00 pm, one resident (R1) requests the AC service to set the temperature to 20\degree C. AC will then try to set the room temperature as requested. Although it may take some time, eventually, AC will set the ambient temperature to 20\degree C. Here, the blue solid curve represents the expected ambient temperature when only AC is working. Another resident (R2) enters the same room and opens the window service at 8:30 pm. Let us assume that the outside temperature is 30\degree C. In this context, hot air may come inside, and it may increase the overall ambient temperature. AC will work hard and gradually set the indoor temperature to 20\degree C. Here, the cyan solid curve represents the ambient temperature when both AC and window are operating simultaneously. Assume that R1 has a temperature preference (i.e., between 18.5\degree C and 22.5\degree C represented by green dotted lines). Ambient temperature may exceed R1's preference range due to window opening by R2. The red zone represents the violating temperature between 8:30 pm and 9:00 pm. In this regard, R1 might experience conflict with R2. 

%\vspace{-1mm}
% \begin{figure}[htbp]
% \center
% \includegraphics[width=\columnwidth]{Images/Motivation.png}
% \vspace{-4mm}
% \caption{Service impact conflict: (a) progressive change, (b) instantaneous change.}
% %\vspace{-1mm}
% \label{motivation}
% \end{figure}

\textit{Scenario 2:}
Suppose two services (light service, window blind service) are located in the living room (Fig. \ref{motivation}(b)). At 8:00 am, R1 requests the light service to set the illumination to 10 lux. Light will then instantly set the room illumination. Let us assume there are no other light sources. Here, the blue solid line represents the expected ambient illumination when only light is working. R2 enters the same room and opens the blind at 8:30 am. Assume that approximately 20 lux light is coming inside through the blind. Thus, it may increase the overall ambient illumination. In this case, the illumination change is instantaneous. Here, the cyan solid line represents the ambient illumination when both light and blind are operating. Assume that R1 has a luminance preference (i.e., between 5 lux and 15 lux, represented by green dotted lines). Ambient illumination may exceed R1's preference range due to blind opening by R2. The red zone represents the violating illumination, and R1 might experience conflict with R2.

Hence, impact conflict is situation-specific and dynamic. Impact conflict detection is challenging due to the following reasons:\looseness=-1

% \vspace{-4mm}
\begin{itemize}[leftmargin=*]
    % \item \emph{Uncertainty}: It refers to unforeseen events or processes that can not be accounted for by the services ahead of time or are unknown to the service but affect its performance.
    \item \textbf{Real-time}: In a smart home, services frequently rely on real-time information for operational decision-making. One of the most common smart home services is an Air conditioning unit. It relies heavily on real-time inside temperatures and outside temperatures to control the ambient temperature.\looseness=-1
    
    \item \textbf{Duration and Scale of Effect}: Any function performed by a service results in an effect, which may vary in duration. Some services require a long time to make small changes (e.g., after opening a window service, ambient temperature increases/decreases progressively), while others require a short time to make significant changes (e.g., after opening a blind service, ambient illumination increases immediately during daytime).
\end{itemize}
% \vspace{-4mm}
Not all the service impacts will lead to conflicting situations. It will depend on residents' preferences. Therefore, it is necessary to quantify the impact and estimate residents' preferences. Only then can true conflict be captured. Capturing true conflicts is essential so that they can be resolved. The objective of smart home services is to provide its residents with convenience. \textit{Therefore, an impact conflict detection framework is required that reflects the services' nature of impact and captures impact conflicts considering residents' preferences}.\looseness=-1

%These conflicts are not considered in these context-aware ambient systems. Some researchers have dealt with multi-user preferences for only media applications in a smart home \cite{nauman2020multimedia, shin2009service}. In reality, a multitude of services and applications coexist in a smart home, such as light service, media applications, TV service, and AC service. A knowledge graph-based conflict model is proposed in \cite{huang2021conflict} to detect different types of IoT service conflicts. However, the proposed approach can only capture conflicts in \emph{single-resident} smart homes.

\section{Definitions and Problem Formulation}
%\vspace{-1mm}

We represent the notion of \textit{IoT Service (S)}, \textit{IoT Service Event (SE)}, \textit{IoT Service Request (SR)}, and \textit{Impact (I)} to explain the concept of \textit{Impact Service Conflict}. The definitions of $S$, $SE$ and $SR$ have been adopted from \cite{chaki2021dynamic}.
% We focus on \textit{shareable} IoT services where conflicts may arise. A \textit{shareable} IoT service serves multiple users at the same time and location. Radio, television, DVD, AC, light, heater, and fan are some examples of shared IoT services. A \textit{non-shareable} IoT service serves only one user at a time. Examples are toaster, microwave oven, electric kettle, and washing machine.

\noindent
\textbf{Definition 1.} An \textit{IoT Service ($S$)} is a tuple of \big \langle \textit{$S_{id}$, $S_{name}$, $F$, $Q$}\big \rangle \hspace{0.15 cm}where:
%\vspace{-2mm}
\begin{itemize}[nosep]
  \item \textit{$S_{id}$} represents the unique service identifier (ID).
  \item \textit{$S_{name}$} is the name of the service.
  \item \textit{$F$} is a set of \big \{\textit{$f_1$,$f_2$,...,$f_n$}\big \} where each $f_i$ is a functional attribute of a service. The purpose of having a service is considered as the function of a service.
  \item \textit{$Q$} is a set of \big \{\textit{$q_1$,$q_2$,...,$q_m$}\big \} where $q_j$ is a non-functional attribute of a service.
\end{itemize}

%\vspace{-1mm}

%For example, a TV service is represented as \big\langle\textit{2, TV, \{telecasting programs, receptor for security camera\}, \{\$1000, 3 years, 500 watts}\big\rangle. 2 is a unique id and TV is the name of a service. Functional properties include telecasting programs via channels (e.g., Fox, Discovery), and a TV may act as a receptor for a security camera. \{\$1000, 3 years, 500 watts\} is a set of non-functional properties like price, warranty, and electricity consumption rate, respectively.

\noindent
\textbf{Definition 2.} An \textit{IoT Service Event ($SE$)} records the service state along with its user, execution time, and location during the service manifestation (i.e., turn on, turn off, increase, decrease, open, close). An \textit{IoT Service Event Sequences ($SES$)} is a set of \big \{\textit{$SE_1$, $SE_2$, $SE_3$,.......$SE_k$}\big\} where each $SE_i$ is a service event. Occupants usually interact with IoT services for various household chores, and the \textit{previous} interactions are recorded as IoT service event sequences. An IoT service event is a tuple of \big \langle \textit{$SE_{id}$, \{$S_{id}, F, Q\}, T, L, U$}\big \rangle \hspace{0.15 cm} where:\looseness=-1

%\vspace{-2mm}

\begin{itemize}[nosep]
  \item \textit{$SE_{id}$} is the unique service event ID.
  \item \textit{$S_{id}$} is a unique ID of the enacted service. \textit{$F$} is a set of functional attributes. \textit{$Q$} is a set of non-functional attributes. 
  %\item \textit{$F$} is a functional attribute of the enacted service.
  %\item \textit{$Q$} is a non-functional attribute of the enacted service.
  \item \textit{$T$} is the time interval of the service consumption. $T$ is a tuple of \big \langle $SET_s, SET_e$\big \rangle \hspace{0.15 cm}where $SET_s$ and $SET_e$ represent the start time and end time of the service.
  %consumption, respectively.
  \item \textit{$L$} is the service event location and \textit{$U$} is user who consumed the service.
\end{itemize}

\noindent
\textbf{Definition 3.} An \textit{IoT Service Request ($SR$)} is an instantiation of a service, and it represents a resident's \textit{current} service requirement. An \textit{IoT Service Request Sequences ($SRS$)} is a set of \big \{\textit{$SR_1$, $SR_2$, $SR_3$,.......$SR_n$}\big\} where each $SR_i$ is an IoT service request. An IoT Service Request ($SR$) is a tuple of \big \langle \textit{$SR_{id}$, \{$S_{id}, F, Q\}, \{SRT_s, SRT_e\}, L, U$}\big \rangle \hspace{0.15 cm} where:

\begin{itemize}[nosep]
  \item \textit{$SR_{id}$} is the unique service request ID.
  \item \textit{$S_{id}$} is a unique ID of the requested service. \textit{$F$} is a functional attribute and \textit{$Q$} is a non-functional attribute of the requested service. $Q$ is a tuple of \big \langle $a, v$\big \rangle \hspace{0.15 cm}where $a$ is the attribute's name and $v$ represents the attribute's value.
  \item \{\textit{$SRT_s$,$SRT_e$}\} represent the requested service's start time and end time.
  \item \textit{$L$} is the location of the service and \textit{$U$} is the user of the service.
\end{itemize}

%\vspace{-2mm}

%An example of an IoT service request is \big\langle \textit{14, \{2, Discovery, 50 dB, \{19:30, 20:30\}, living room, U1}\big\rangle. 14 is the ID of this service request and 2 is the ID of the TV. The user wants to watch the TV between 19:30 and 20:30 on the day when the request is made. Discovery and 50 dB are the values of functional and non-functional attributes such as channel and volume of TV. Living room and U1 denote the location and user of the requested service, respectively.

\noindent
\textbf{Definition 4.} \textit{Impact ($I$)} specifies the effect of simultaneous service requests on the ambient environment properties (i.e., temperature, illumination, sound, humidity). Suppose that two requests are placed to two completely different services, but they may interfere with each other via an environment property. For instance, one resident wants a cool ambient environment by using AC service, and another resident wants fresh air by using window service. If the outside temperature is higher than the inside temperature, hot air may come inside and impacts the ambient temperature. Formally, given two users' service requests ($SR_i$ and $SR_j$) at time $t$, an impact ($I$) may occur if $\exists qi_a = qj_a$, such that $qi_v \neq qj_v$. It is represented: 

\begin{equation}
    I(SR_i, SR_j) = <\{S, Q\}, t, val>
    \label{impact}
    %\vspace{-1mm}
\end{equation}
where $\{S, Q\}$ is the impacted service and its associated attribute, $t$ is the time in which the impact may occur, and $val$ represents the severity of the impact.

\noindent
\textbf{Definition 5.} \textit{Impact Conflict ($IC$)} in IoT environment is heavily dependent on users' preferences. Not all the impacts will lead to a conflicting situation. Sometimes a small impact may create a conflict, whereas sometimes, a big impact may not create a conflict. On the one hand, assuming that a person has a very strict sound preference, a small change in the ambient sound may create a conflict for them. On the other hand, assuming that the same person has a very flexible temperature preference, a big temperature change may not create a conflict. Note that change in ambient environment properties is caused by another user's service enactment and is captured by the impact ($I$) model.

% \vspace{-4mm}

\subsection*{Formal Problem Statement:}

An IoT service ($S$) is associated with a set of functional and non-functional properties. An IoT service event ($SE$) illustrates a resident's previous service usage in conjunction with time and location. IoT service event sequences ($SES$) record all the history of service events, and preferences ($Pref$) can be estimated from these previous events. An IoT service request ($SR$) captures a resident's current service usage requirement. Multiple residents' requirements are stored in service request sequences ($SRS$). Impact ($I$) may occur on the ambient environment properties due to the different service requests from residents. Consequently, an impact conflict ($IC$) may happen if a change in ambient properties exceeds any user's preference range. Given this information, the paper aims to identify a function $F(S, SRS, SES)$, where $IC \approx F(S, SRS, SES)$. It is represented as:

\begin{equation}
    IC(S, SRS, SES) = <\{S, Q\}, T, L, U, P>
    \label{impact_conflict}
    %\vspace{-1mm}
\end{equation}
where $\{S, Q\}$ is the impacted service and its associated attribute, $U$ is the user who experiences the conflict, $P$ is the likelihood of the conflict, and $T$ and $L$ represent the time and location, respectively, of the probable impact conflict.

\section{Impact Conflict Detection Framework}

\begin{figure*}%
    \centering
    \subfloat[\centering System architecture ]{{\includegraphics[width=0.4\textwidth]{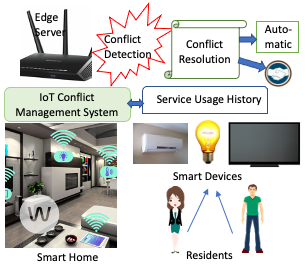} }}%
    \qquad
    \subfloat[\centering Proposed framework ]{{\includegraphics[width=0.4\textwidth]{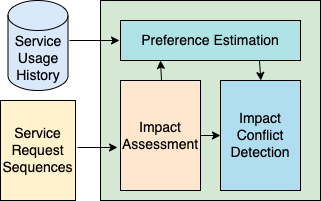} }}%
    %\vspace{-1mm}
    \caption{System architecture and conflict detection framework.}%
    % \vspace{-8mm}
    \label{systemmodel}%
\end{figure*}

Fig. \ref{systemmodel}(a) illustrates an abstract view of the conflict management system, and Fig. \ref{systemmodel}(b) shows our proposed framework. Note that conflict resolution is out of the scope of this paper. However, we keep it in mind when designing our framework. The framework has 3 modules: (i) impact assessment, (ii) preference estimation, and (iii) conflict detection. The impact is quantified from the residents' present service requirements. Preference is estimated from the residents' previous service usage. Impact and preferences are then fed into the conflict detection module to estimate the likelihood of conflicts.\looseness=-1

% \vspace{-3mm}

% \vspace{-4mm}

\subsection{Impact Assessment:}
% \vspace{-3mm}

In this module, we quantify the impact from the users' service requests. The proposed impact model is designed considering four ambient environment properties such as temperature, illumination/brightness, sound, and humidity. Two completely different services may interfere with each other via these environment properties due to their simultaneous execution. The impact may occur based on the change of an environmental variable; either it can be a progressive change or an instantaneous change. Given two service requests ($SR_i$, $SR_j$), if the following conditions are satisfied, only then the impact is computed.

%Due to lack of space, we only provide the definitions of different types of conflicts.
\begin{itemize}[nosep]
    \item $L_{S_i}$ $\simeq$ $L_{S_j}$, meaning, two services ($S_i$, $S_j$) are executed at the same location.
    \item $(SRT_{si}, SRT_{ei}) \cap (SRT_{sj}, SRT_{ej})) \neq \emptyset$, denoting that two service requests ($SR_i$, $SR_j$) are invoked at the same time, and there is a temporal overlap.
    \item $U_{S_i} \neq U_{S_j}$ and $S_i \neq S_j$ , meaning these two requests are invoked on two different services by two different users.
    \item $\exists Q_k \in S.Q: S_i.Q_k \simeq S_j.Q_k$; there exists at least one property which is similar between $S_i.Q$ and $S_j.Q$ (e.g., temperature linked with AC and window).
\end{itemize}

At first, we find the overlapping segment ($OS$) (i.e., intersection) between two service requests. Starting time ($OST$) of $OS$ would be the maximum of the two start times in $SR_i$ and $SR_j$. Ending time ($OET$) of $OS$ would be the minimum of the two end times in $SR_i$ and $SR_j$ (e.g., for two requests, if two ranges are 8 pm to 10 pm and 9 pm to 11 pm, the $OS$ is 9 pm to 10 pm). We use Signal Temporal Logic (STL) based runtime verification technique to measure the degree of impact. STL is a formalism used to specify real-time properties of discrete (e.g., instantaneous change) and continuous (e.g., progressive change) signals (e.g., illumination, temperature). We refer readers to \cite{donze2010robust} for the formal definition of STL semantics. We find that most smart home service requirements can be specified using STL formula in the form of $\square_{(a, b)} (x := \lambda)$ where $x$ is a signal about environment property and $\lambda$ is a setpoint (i.e., threshold). For instance, considering motivation scenario 1, STL formula for resident R1's service requirement, $SR_1$: $\square_{(8:00, 9:00)} (temperature := 20\degree C)$. We now describe how to compute the degree of impact in which a signal, either continuous or discrete, about an environment property violates a service requirement. A notion of robustness value for violating STL formulas is formally defined in \cite{donze2010robust}. The robustness value of a signal $x$ violating $(a,b) (x:=\lambda)$ at time $\tau$ is defined as:

% \vspace{-6mm}
\begin{equation}
% \vspace{-2mm}
    \rho = sup_{t \in (\tau+a,\tau+b)} (x(t)-\lambda)
    \label{robustness}
    %\vspace{-1mm}
\end{equation}

\begin{figure*}[htbp]
\center
\includegraphics[width=\textwidth]{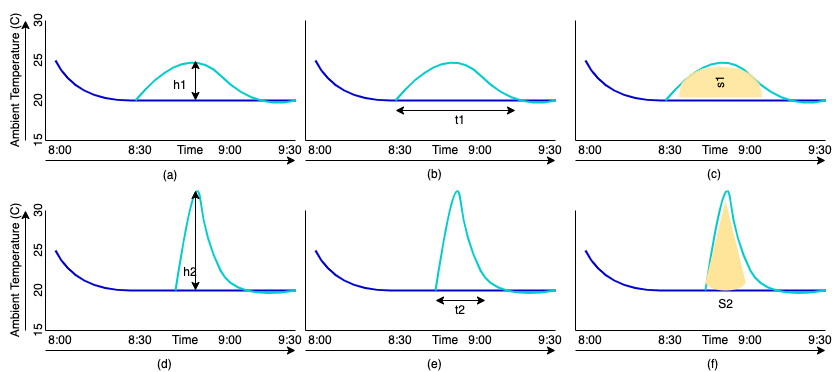}
%\vspace{-4mm}
\caption{Impact assessment: (a \& d) robustness measurement, (b \& e) percentage of impact time, (c \& f) integral of deviation.}
%\vspace{-6mm}
\label{impactmodel}
\end{figure*}

Intuitively, the robustness value indicates extremum points of the signal. The robustness value is useful for telling the worst-case performance, but it does not show the average or overall performance. For impact assessment, we are interested to know both. We use the following example to describe why measuring the robustness value only is not enough for finding the severity of impact. Recall our motivation scenario 1. The impact may occur between 8:30 pm and 9:00 pm. We compute impact considering R1's requirement $SR_1$: $\square_{(8:00, 9:00)} (temperature := 20\degree C)$. In this case, the temperature is between 20\degree C and 25\degree C for 30 minutes. Therefore, the robustness value ($h1$) is 5 (Fig. \ref{impactmodel}(a)). Assume that, in another case, the outside temperature is 40\degree C, and the AC has a higher capacity to cool. In this context, the temperature fluctuates between 20\degree C and 30\degree C for 10 minutes. The robustness value ($h2$) is 10 (Fig. \ref{impactmodel}(d)). If only robustness is considered for impact assessment, then the second case will have a higher impact. Thus, they will have a higher conflict likelihood. However, this is not always true since the impact duration is very short. Although the first case has lower robustness, it may have a higher likelihood of conflict as the impact duration is long.

To address this limitation, we present a concept of signal deviation to measure the degree of impact. At first, we calculate the percentage of time when a requirement is violated. To start with, we define Equations \ref{positivetimepercentage} and \ref{negativetimepercentage} to calculate the positive part and the negative part of a function f(x), respectively.

% \vspace{-4mm}
\begin{equation}
\label{positivetimepercentage}
%\vspace{-2mm}
    \theta^+(x) = max(f(x), 0) = 
    \begin{cases}
      f(x), & \text{if }  f(x) > 0 \\
      0, & \text{otherwise}
    \end{cases}
    %\vspace{-1mm}
\end{equation}

% \vspace{-4mm}
\begin{equation}
\label{negativetimepercentage}
%\vspace{-2mm}
    \theta^-(x) = min(f(x), 0) = 
    \begin{cases}
      f(x), & \text{if } f(x) < 0 \\
      0, & \text{otherwise}
    \end{cases}
    % \vspace{-1mm}
\end{equation}

We compute the percentage of time given a signal $\square_{(a, b)} (x := \lambda)$ as:

% \vspace{-3mm}
\begin{equation}
%\vspace{-2mm}
    \eta = \frac{1}{b-a}\int_{\tau+b}^{\tau+b}sgn(|\theta^-(x(t)-\lambda)|)dt
    \label{timepercentage}
    % \vspace{-1mm}
\end{equation}

We then use Equation \ref{integral} to compute the integral of signal deviations accumulated in a period when the requirement $\square_{(a, b)} (x := \lambda)$ is violated.

% \vspace{-3mm}
\begin{equation}
%\vspace{-2mm}
    I = \eta \int_{\tau+b}^{\tau+b}(|\theta^-(x(t)-\lambda)|)dt
    \label{integral}
    % \vspace{-1mm}
\end{equation}

% \vspace{-5mm}

\subsection{Preference Estimation:}
% \vspace{-2mm}
In this module, we estimate users' preferences for a service based on previous usage records. Impact module outputs the degree of impact, the name of the impacted services, and their overlapping time period where an impact may occur. We use overlapping impacted time periods to extract residents' service usage patterns from their previous history. Then, it calculates the preference range of frequently used services. We scan the previous history to find out all the overlapping service events (algorithm 1). The input of this algorithm is the previous service usage dataset ($DB$) and the overlapping segment ($OS$) when an impact may occur. All the previous events that have overlapped with the given overlapping segment are the output of this algorithm. Each location has IoT services, and we mine out the frequent service interactions based on location. We can determine if a person uses specific services more frequently by evaluating their service interactions over a long period. A distance-based clustering algorithm, DBSCAN, is employed, resulting in clusters containing services that the resident has used many times during this overlapping period \cite{choksi2022you}. For example, an impact related to a light service occurs in the living room between 20:00 and 20:30. This component searches all the service events that previously occurred, either partially or fully, between 20:00 and 20:30 in the living room and stores them in a list ($OSE$). This list contains the overlapping service events and their timestamps.

% \vspace{-5mm}

\setlength{\textfloatsep}{0pt} % For vertically reducing space under algorithm.
\begin{algorithm}[t!]
\small
\caption{Overlapping Service Events}\label{alg:algorithm1}
\begin{algorithmic}[1]
\REQUIRE
$DB$, $[OST,OET]$ // impacted time-period $[s,e]$
\ENSURE
$OSE$ // overlapping service events along with time interval

\STATE {$TM = \emptyset, OSE = \emptyset$}

% \item[] // Clustering services based on location
% \FOR{\textbf{each} $se_i$ in $DB$}
% \FOR{\textbf{each} $l_j$ in $se.l$}
% \FOR{\textbf{each} $s_k$ in $se.S$}
% \IF{$s_k.l$ is equal to $l_j$}
% \STATE $LSE_i \leftarrow insert(s_k)$
% \ENDIF
% %\STATE $us_i \leftarrow unique(services)$
% \ENDFOR
% \ENDFOR
% \ENDFOR
\item[] // Finding overlapping service events
%\textbf{Step 2. Find Common Services}
\FOR{\textbf{each} $se_i$ in $DB$}
\FOR{\textbf{each} $s_j$ in $se_i$}
%\STATE $CS \leftarrow count(s_j)$ 
\IF{$s_j.L == se_i.L$}
\IF{$s_j.SET_s$ or $s_j.SET_e$ falls between $[OST,OET]$}
\STATE $TM \leftarrow addTimeInterval(s_j.SET_s, s_j.SET_e)$ 
\STATE $OSE \leftarrow insert(s_j, TM)$
\ENDIF
\ENDIF
\ENDFOR
\ENDFOR
%\STATE $SV \leftarrow sort(TM)$
%\STATE $SS \leftarrow top (SV,k)$
%\STATE $TI \leftarrow timeInterval(S)$
%\STATE $OSE \leftarrow overlap(SV)$
\RETURN $OSE$

\end{algorithmic}
\end{algorithm}

\begin{figure}[t!]
\begin{center}
%\vspace{-2mm}
\includegraphics[width=\columnwidth]{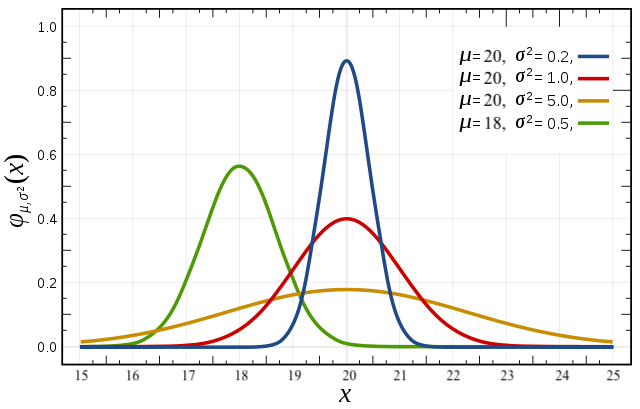}
% \vspace{-2mm}
\caption{Temperature distribution to get optimal preference range.}
% \vspace{-3mm}
\label{probDist}
\end{center}
\end{figure}

In order to estimate preferences from this list, we then plot all the data points in a two-dimensional plane. Note that, in this paper, we only deal with numerical data associated with IoT services such as temperature, illumination, and sound. X-axis represents each service attribute's values (e.g., AC temperature, light illumination), and Y-axis represents frequency. We extract the range from the X-axis, where most data points are located. We get the optimal preference range (a.k.a. preference band) based on a dynamic programming approach from the plotted values. The width of the preference band can be of different sizes. For example, Fig.\ref{probDist} depicts four residents' preferred AC temperature range. Each color denotes each resident's preferred temperature distribution. The blue curve represents a very strict (i.e., very narrow range of AC temperature) nature of a resident. In contrast, the yellow curve represents another resident's very flexible (i.e., very wide range of AC temperature) nature. This depicts the residents' actual preferences in terms of range.\looseness=-1

% \vspace{5mm}

\subsection{Service Impact Conflict Detection}
% \vspace{-3mm}

We use the preference range generated from the preference estimation module and the impact weight generated from the impact assessment module to estimate the conflict likelihood. \textit{IoT Service Impact Conflict} occurs when two services cannot satisfy the preferences of one or multiple users at the same location and time duration. If there is no impact (i.e., the value of impact weight is 0), we assume there is no impact conflict. For example, one resident prefers to have dim light while watching movies in a theatre-like environment. Another resident enters the same room at the same time and opens the window blind to enjoy the outside beauty (assuming the window is glass-made). If it is night and no light enters, there is no impact of opening the blind; thus, there is no conflict. If the value of the impact score is greater than 0, we only then check the preference range of the residents and the change in the ambient environment. 

At first, we follow a preferential proximity approach to find the likelihood of having a conflict. Suppose we have two ranges of ambient temperature. One range denotes the resident's preferred temperature captured from the preference estimation module. Assume one resident's preference is between 19\degree C and 21\degree C, represented as [a\_s, b\_s] where a\_s = 19 and b\_s = 21. Another range denotes fluctuated ambient temperatures captured from the impact assessment module. Assume it is between 20\degree C and 23\degree C and is represented as [a\_e, b\_e] where a\_e = 20 and b\_e = 23. Preferential proximity is computed as follows:

% \vspace{-3mm}
\begin{equation}
%\vspace{-1mm}
    Pref_{prox} = \frac{|Median(a_s, a_e, b_e) - Median(b_s, a_e, b_e)|}{Max(b_s, b_e) - Min(a_s, a_e)}
    \label{prefProx}
    %\vspace{-1mm}
\end{equation}

Here, the value of $(1-Pref_{prox})$ determines the likelihood of conflicts. For the example mentioned above, the value of $Pref_{prox}$ is 0.25. Therefore, the likelihood of conflict is (1-0.25) = 0.75. If the value of $Pref_{prox}$ becomes 1, then the preferred range and fluctuated range match completely, and no conflict will occur. However, if the value is 0, then there is no range overlapping, and there will be a high likelihood of conflicts occurring. The more the preferential proximity value leans towards 0, the more possibility of having a conflict. Then, we use the temporal proximity strategy to find out the weight of the potential conflict time over impacted time. For instance, the impact overlapping segment is [8:30 pm - 9:20 pm], and the potential conflict time is [8:45 pm - 9:00 pm] (see motivation scenario 1). The temporal proximity technique for evaluating the distance between time-interval data is adopted from \cite{shao2016clustering}. For each impact overlapping segment, $OS_i$, we use a function $f_i$ with respect to $t$ to map the temporal aspect of $OS_i$. Segment start time and end time are represented with $OST_i$ and $OET_i$, respectively. $f_i$ is formalized in Equation \eqref{proximity}.

% \vspace{-3mm}

\begin{equation}
\label{proximity}
%\vspace{-2mm}
    f_i(t) = 
    \begin{cases}
      1, & t \in [OST_i, OET_i] \\
      0, & otherwise
    \end{cases}
    % \vspace{-1mm}
\end{equation}

We generate a set of functions ${f_1, f_2, ... f_n}$ corresponding to the overlapping segments ($OS$) and potential conflict time. Equation \eqref{temporal proximity} calculates the temporal proximity ($temp_{prox}$) for all the overlapping events.

% \vspace{-4mm}

\begin{equation}
%\vspace{-2mm}
    temp_{prox} = \frac{\int_{t_1}^{t_{2n}}\sum_{i=1}^{n}f_i(t)dt}{(t_{2n}-t_1).n}
    \label{temporal proximity}
    %\vspace{-1mm}
\end{equation}

Here, $t_1$ and $t_{2n}$ are the first and last time information of overlapped events from $OS$, and $n$ is the number of instances. Consider the following two events of potential impact conflict. One impacted segment is between 20:00 and 21:00; a potential conflict time is between 20:45 and 21:45. Here, potential conflict time denotes a time-period when the fluctuated signal may exceed any user's preference range. Using Equation \ref{temporal proximity}, the temporal proximity of these two intervals can be calculated as ((20:45-20:00)+(21:00-20:45)*2+(21:45-21:00))/((21:45-20:00)*2)\newline=0.57. Consider another impacted segment is between 18:00 and 19:00. An potential conflict time is between 18:10 and 19:10. The temporal proximity of these two events can be calculated as ((18:10-18:00)+(19:00-18:10)*2+(19:00-19:10))/((19:10-18:00)*2)=0.86. Thus the latter case has more weight while calculating the likelihood of a conflict. Therefore, for each pair of overlapping service requests, conflict likelihood ($CL$) is computed, including impact weight ($I$), preferential proximity ($Pref_{prox}$), and temporal proximity (($temp_{prox}$)) as:

% \vspace{-3mm}
\begin{equation}
%\vspace{-2mm}
    CL = I * ((1-Pref_{prox}) + temp_{prox})
    \label{conflictlikelihood}
    % \vspace{-3mm}
\end{equation}

\section{Experimental Results and Discussion}
% \vspace{-5mm}
\subsection{Dataset and Experimental Setup}
% \vspace{-3mm}
We use a dataset collected from the Center for Advanced Studies in Adaptive Systems (CASAS) to evaluate the proposed conflict detection framework \cite{cook2012casas}. There exist a few multi-resident activity datasets. Nevertheless, these datasets do not capture conflicting situations for service usage. Thus they are not helpful enough for our experiment. These publicly available datasets reflect the compromises of the residents interacting with services. As a result, they do not have any IoT service conflicts. Records of a single resident interacting with IoT services, on the other hand, reveal an individual's true preferences. Consequently, we combine individual inhabitants' service interaction records (labels HH102, HH104, HH105, and HH106 are used from the CASAS dataset) to mimic a multi-resident smart home environment. These labels were chosen because they feature actions from the same time period (between June 15, 2011, and August 14, 2011). Table \ref{dataset} contains descriptions of dataset attributes.

%The dataset consists of different sensors, such as temperature, light, infrared motion, magnetic door, infrared motion, and battery level. In this research, we consider each sensor as an IoT service.

%\vspace{4mm}

\begin{table}[htbp]
% \vspace{-3mm}
\caption{Description of the dataset attributes}
% \vspace{-4mm}
\label{dataset}
\center
\begin{tabular}{|c|l|p{4.5cm}}
\hline
\textbf{Attributes} &  \textbf{Description} \\
\hline
Date &  The service execution date\\ \hline
Time &  The service execution time\\ \hline
Sensor & \parbox{6.5cm}{Name of the sensors such as motion sensors, light switch, light sensors, door sensors, temperature sensors}\\\hline
Status & \parbox{6.5cm} {ON, when the service starts, and OFF, when the service stops}\\
\hline
\end{tabular}
%\vspace{-2mm}
\end{table}

%Sensor's on/off time is included in the dataset. The terms ``ON" and ``OFF" refer to the service's start time and end time, respectively. The light and temperature sensors gathered light illumination levels and temperature values in the dataset. We consider these values as the inhabitant's preferred values related to light and AC service. We retrieve outside temperature from an online database\footnote{https://www.worldweatheronline.com/}.

% \vspace{-5mm}
\subsection{Ground-truth and Metrics}
% \vspace{-3mm}
The service interactions from various users overlap in terms of time in the merged dataset. The dataset is annotated with different activity labels and has the layout of the home. However, it lacks interaction records related to windows and blinds. Therefore, we create events such as opening and/or closing window and blind services to mimic the realistic environment. Each room has a thermostat, and we assume that the set temperature is the preferred temperature for a user. We then estimate how much time it will take to cool/hot inside temperature if the window is opened using this formula, ``Time taken to cool/hot (in hours) = (volume of the room * density of air * change in temperature * specific heat of air)/ (latent heat of fusion * ton capacity of AC * 1000 / 24). For simplicity, we keep most of the variables constant. The dataset has ``Watch\_TV" activity label and indoor light level. However, the sound volume and outdoor illumination information are missing. Hence, we augment the dataset by randomly generating these values based on a uniform distribution. After that, whenever two users simultaneously use two services that set the state of an environment property in contradictory ways, a conflict observation is created. We assume that a conflict involving more than two parties can be broken down into numerous pairwise conflicts. We discretize numerical contexts to produce potential context scenarios, and then we tally the occurrences of each scenario in the data; this result reflects how frequently a user has encountered a certain context. If a conflict observation is made at each occurrence, the number of conflicts is increased. Each individual user and service pair's count is kept individually. Thus, the number of conflicts experienced by that pair of users for that environment property divided by the number of times that context scenario occurred is used to calculate the ground-truth likelihood for two users to have conflicts for a service under a possible scenario. We first evaluate our approach based on accuracy, precision, and recall metrics. Then, we represent the performance across the studies using the mean absolute error (MAE). We can contrast the likelihood based on the ground truth with the likelihood predicted by the method. For instance, to calculate the anticipated likelihood for each sample while evaluating our methodology, we look up the likelihood in the conflict scenario that includes this sample. For a set of samples, 
the number of samples, the $MAE = \frac{1}{N} \sum|ol-el|$ where $ol$ is ground-truth likelihood and $el$ is the estimated likelihood of the proposed approach.

%Each sample is linked to a pair of users and services that indicate the actual possibility of any disputes between the two users regarding a given environmental attribute.

% \vspace{-5mm}
\subsection{Performance Evaluation}
% \vspace{-3mm}

\begin{figure*}[t]%
    \centering
    \subfloat[\centering Without preference ]{{\includegraphics[width=0.4\textwidth]{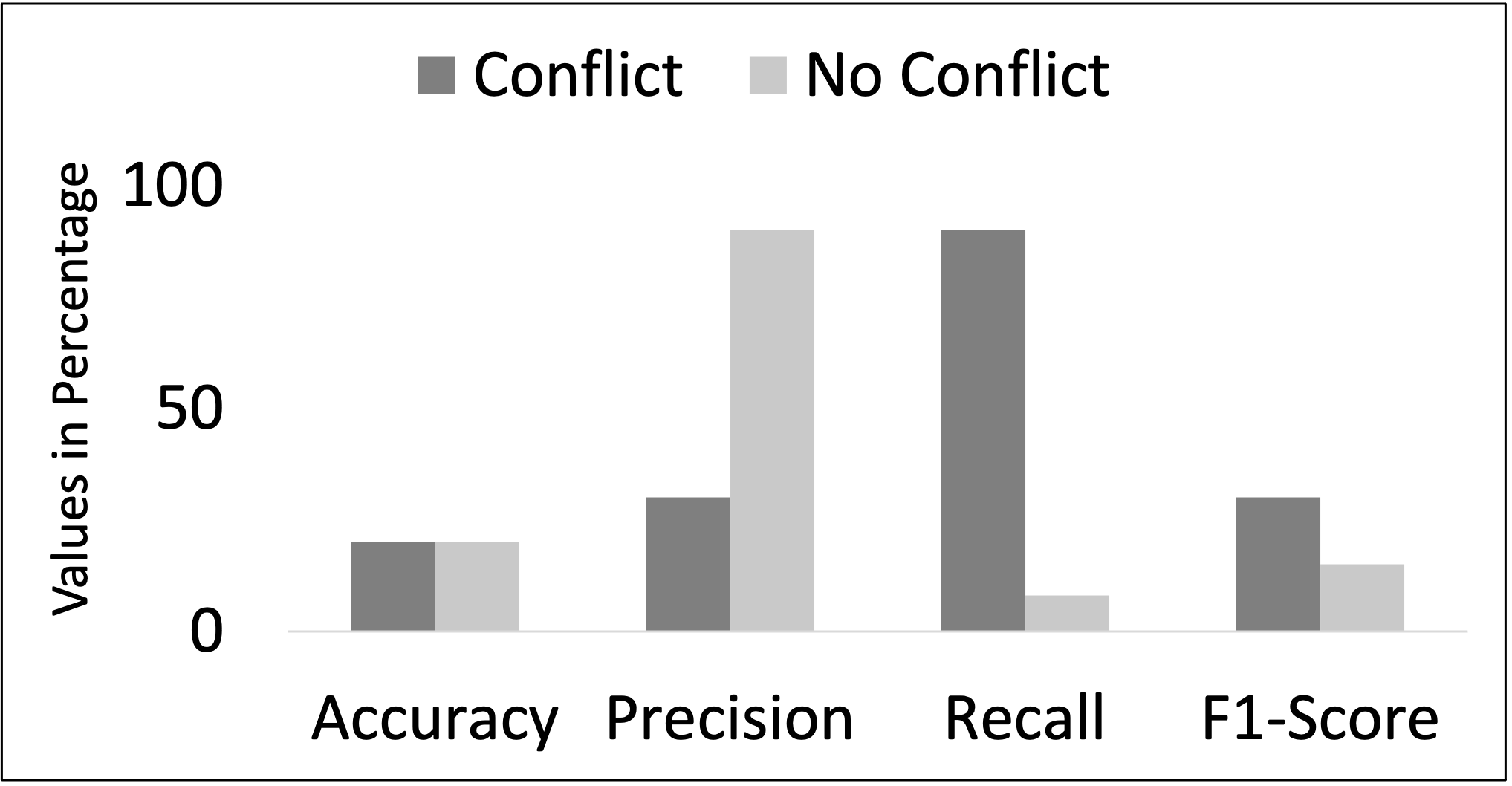} }}%
    \qquad
    \subfloat[\centering With preference ]{{\includegraphics[width=0.4\textwidth]{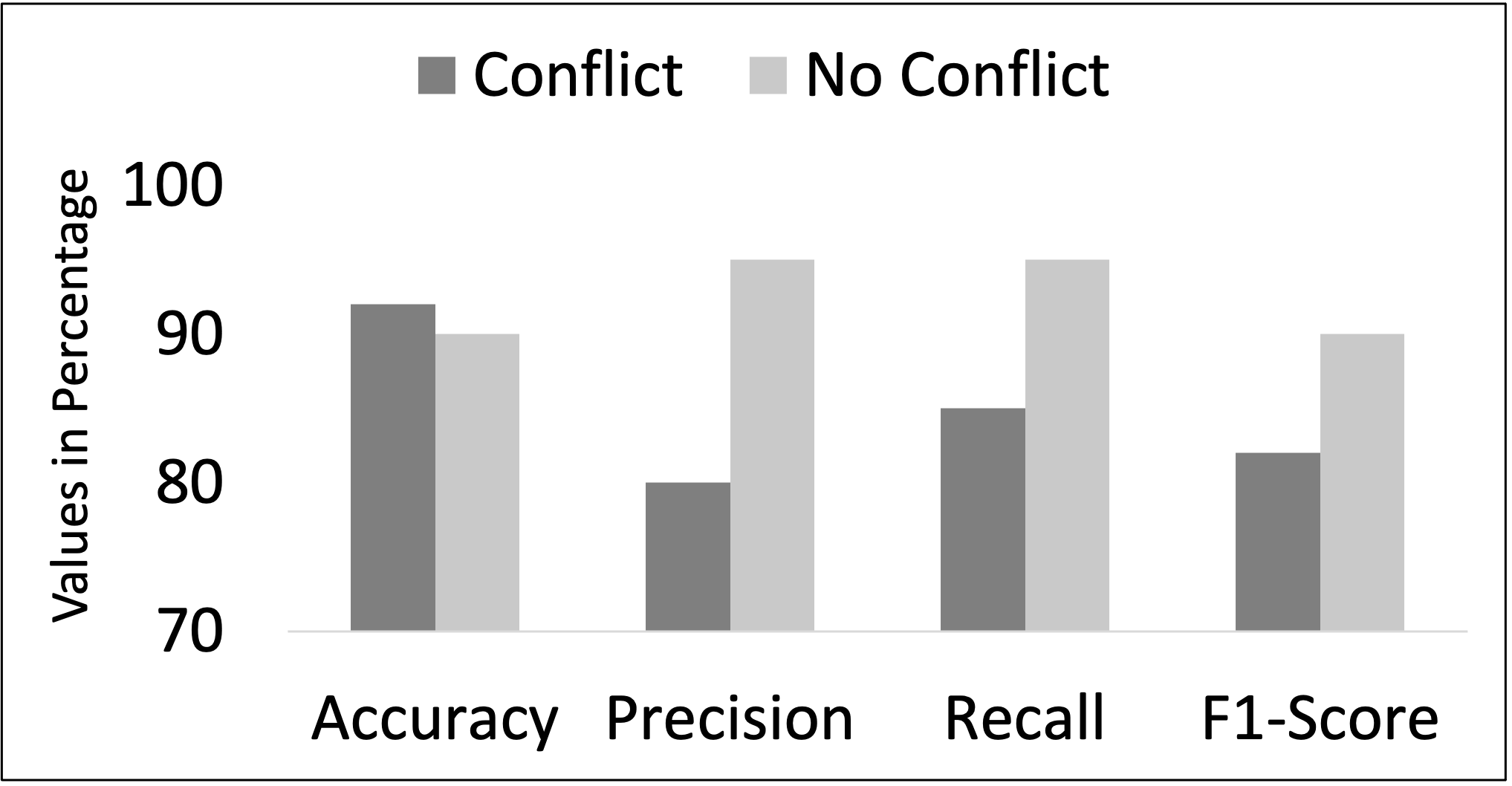} }}%
    %\vspace{-1mm}
    \caption{Experimental results based on accuracy, precision, recall, f1-score.}%
    \vspace{2mm}
    \label{results}%
\end{figure*}

At first, we measure our proposed framework's accuracy, precision, and recall score. The first set of experiments is conducted without considering residents' preferences (fig. \ref{results}(a)). This set only considers impact as a binary decision (i.e., if there is an impact, there is a conflict). It is visible from the figure that the accuracy score is very low when preference is not considered. Because this approach detects a scenario as a conflict even when the temperature difference is only 1\degree C as 1\degree is considered as an impact here. It counts many non-conflicting situations as conflicts, thus increasing false positive numbers. This reduces the approach's overall accuracy. However, this approach performs better while detecting non-conflicting situations. This is why the precision and recall scores regarding no-conflict are high. Then, we conduct the second set of experiments considering preference scores. This approach detects both conflicting and non-conflicting situations more accurately based on precision, recall, and f1-score (fig. \ref{results}(b)). This approach considers preference as a range. Therefore, when an impact occurs, it checks the residents' interaction history and mines out the preference range. If the impact value falls within the preference range, then it does not detect the situation as a conflict. If the impact value falls beyond the preference range, only then is it accounted as conflict. Our proposed approach achieves a high accuracy score of about 90\%. Precision, recall, and f1-scores are also high.

% 95\%, 95\%, and 90\%, respectively, for detecting no-conflicts. The precision score for conflict detection is moderate (about 80\%) as it sometimes fail to capture residents' strict nature of preferences.

\begin{figure*}[t]%
    \centering
    \subfloat[\centering Temporal proximity threshold ]{{\includegraphics[width=0.4\textwidth]{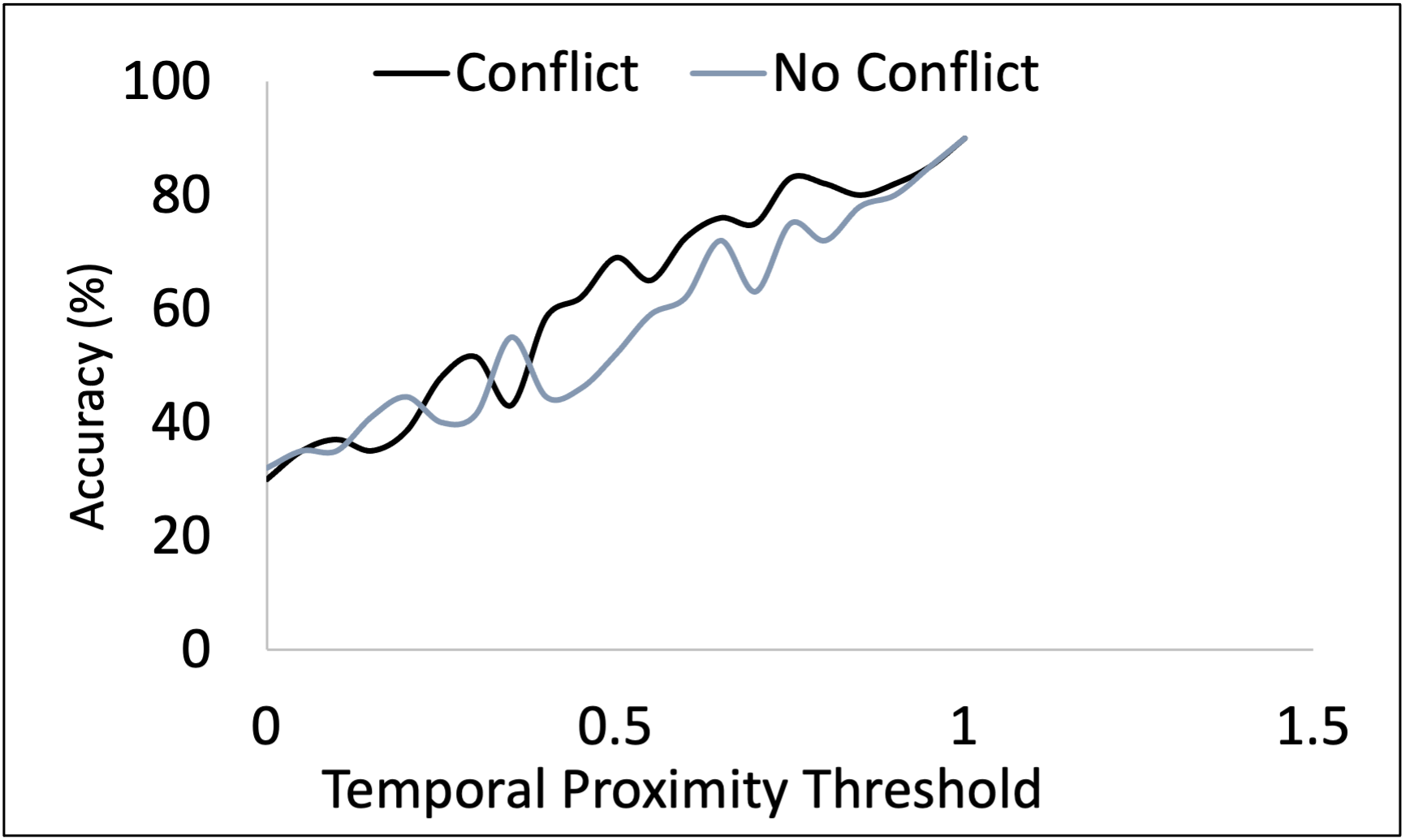} }}%
    \qquad
    \subfloat[\centering Preferential proximity threshold ]{{\includegraphics[width=0.4\textwidth]{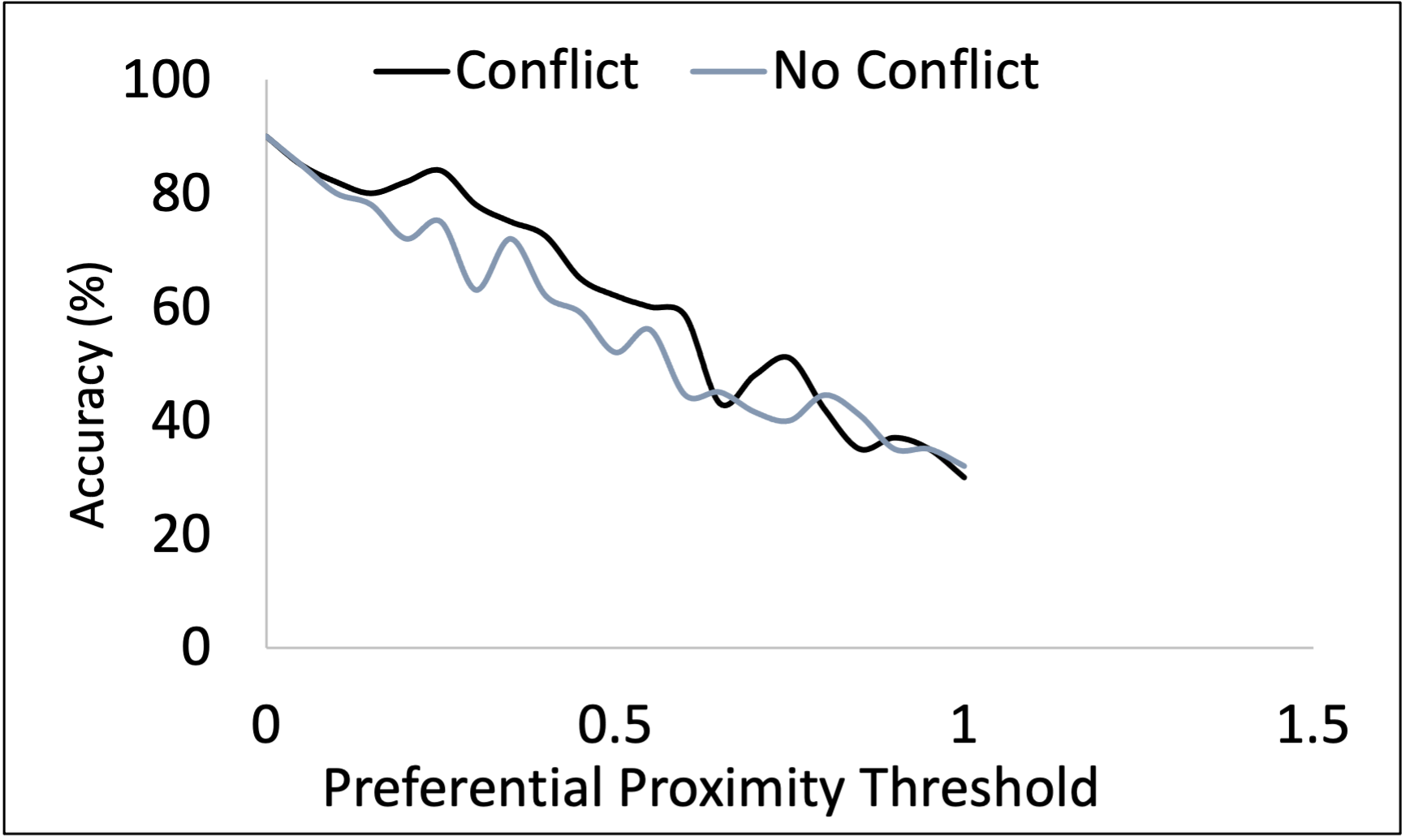} }}%
    %\vspace{-1mm}
    \caption{Change in accuracy with different proximity threshold.}%
    % \vspace{2mm}
    \label{threshold}%
\end{figure*}

We conduct the third set of experiments to evaluate the conflict likelihood value based on MAE. The proposed approach performs well regarding temperature and illumination variables (Table \ref{mae}). The reason is our impact model can successfully capture the gradual change and instantaneous change with time granularity. It also captures users' preference ranges, taking into account different contextual factors. However, our model does not perform well in detecting sound impact conflict. Because the proposed preference estimation model cannot extract residents' preferences when no service request data is available. For example, a resident prefers to have less sound as they want to study. Here, the study is an activity, and there is no sound data associated with service requests. Finally, we adjust the temporal and preferential proximity values to observe the detection accuracy variation. Fig. \ref{threshold} demonstrates that increasing temporal and decreasing preferential proximity capture conflicts more accurately. When the temporal threshold is small, and the preferential threshold is big, there are lots of overlapping events that are pruned. However, increasing the temporal and decreasing the preferential threshold denotes more capturing ability of conflicts.
%This is why, when the threshold is 1.0, no events are discarded and all the conflicts are detected (100\%) like baseline approach.

%Therefore, our model cannot capture this type of preferences properly. Hence, likelihood estimation is preferred when detecting impact conflicts for real-world situations.

% \vspace{-2mm}
\begin{table}[htbp]
%\vspace{-3mm}
\caption{Average Mean Absolute Error for Environment Variables}
% \vspace{-4mm}
\label{mae}
\center
\begin{tabular}{|c|c|c|c|}
\hline
\textbf{Environment Property} & \textbf{Conflict} & \textbf{No Conflict} &  \textbf{Overall}\\
\hline
Temperature & 9.3 $\pm$ 0.3 & 1.2 $\pm$ 0.2 & 4.6 $\pm$ 0.2 \\ \hline
Illumination & 13.1 $\pm$ 1.4 & 24.4 $\pm$ 6.0 & 13.8 $\pm$ 1.7 \\ \hline
Sound & 24.4 $\pm$ 0.9 & 0.5 $\pm$ 0.0 & 1.0 $\pm$ 0.0 \\
\hline
\end{tabular}
%\vspace{-3mm}
\end{table}

% \vspace{-6mm}
\section{Related Work}

The exploration of conflict detection spans across various domains, notably in software engineering and smart home environments. They reflect diverse challenges and methodologies tailored to each field's unique requirements. In software engineering, the focus is on reconciling stakeholder requirements, while in smart homes, conflicts arise from the interactions among devices/services and occupants, necessitating innovative approaches to ensure seamless cohabitation and system functionality. This section delves into the existing literature to understand the mechanisms of conflict detection, highlighting the differences and complexities inherent to these domains.

\subsection{Conflicts in Software Engineering}

Conflict detection and resolution are key in software engineering, especially during requirement engineering, which involves defining a system's needs and limitations. Differences between managing conflicts in software engineering versus smart home IoT environments highlight the unique challenges in each domain:

\begin{itemize}[leftmargin=*]

\item \textbf{Nature of conflicts:} Software engineering conflicts often emerge from stakeholders' differing needs during the requirements phase, usually due to misinterpretations or inconsistent priorities \cite{easterbrook1994resolving,nuseibeh1994framework}. In smart homes, conflicts arise from interactions among multiple devices, services, and users, leading to issues like resource contention or privacy concerns \cite{gubbi2013internet,perera2013context}.

\item \textbf{System complexity:} Smart homes involve complex, interconnected physical systems influenced by dynamic factors, necessitating consideration of physical constraints and safety \cite{gubbi2013internet}. Conversely, software engineering conflicts focus more on aligning stakeholder expectations without the immediate impact of real-time physical interactions \cite{easterbrook1994resolving,nuseibeh1994framework}.\looseness=-1

\item \textbf{Stakeholders:} While software engineering conflicts mainly involve human stakeholders (i.e., developers, managers, customers), smart home conflicts also include interactions between IoT devices and services, requiring a more comprehensive approach to conflict management \cite{perera2013context}.\looseness=-1

\end{itemize}

\subsection{Conflicts in Smart Homes}

``Conflict is a natural disagreement between different attitudes, beliefs, values, or needs'' \cite{wang2011development}. A home in Australia, on average, has 17 connected devices (i.e., services)\footnote{https://tinyurl.com/vg3lh6w}. In addition, 2.6 residents usually live per household\footnote{https://tinyurl.com/y6jzjwf4}. \textit{Therefore, there is a high chance of conflict in multi-occupant homes.} The concepts of conflict and conflict detection are surveyed in the relevant literature. We identify three criteria to categorize conflicts: (i) source, (ii) intervenience, and (iii) solvability. Conflicts may occur depending on different types of sources \cite{ibrhim2021conflicts}. A conflict may occur when many users utilize a resource such as a TV defined as a \textit{resource-level conflict} \cite{lalanda2017conflict}. A conflict may occur when several applications try to use a resource simultaneously. For example, \emph{application-level conflicts} involve building management applications and a human trying to control a light bulb inside a room \cite{tuttlies2007comity}. A conflict may emerge between space and the user when a user disputes space policies. For instance, a user's smartphone may ring inside a library with a silence policy \cite{hua2019riot}. A conflict may also occur when diverse user inclinations in a similar setting are known as \emph{profile-level} conflict \cite{carreira2014towards}. For instance, one user likes to enjoy a TV program with the room light at half limit, and another user wants to peruse with the lights at the full limit. Time is another crucial factor for conflict detection \cite{miandashti2020empirical}.

Intervenience is a key criterion that may cause conflicts \cite{yagita2015application}. Conflict is usually common in multi-resident smart homes, yet a conflict may occur in a single-resident home. For instance, a conflict may occur based on contradictory intentions such as comfort and saving energy simultaneously. An object-oriented approach is proposed in \cite{nakamura2005feature} to detect conflict in a single-occupant smart home. The model is further improved in \cite{nakamura2013considering} by considering environmental requirements and impacts. The proposed effect model is qualitative (i.e., either an effect exists or it does not exist at all), which is the limitation of their work. Solvability is another key criterion for conflict categorization \cite{goynugur2016automatic}. Usually, a conflict is detected during its occurrence (i.e., run-time) \cite{huang2021conflict}. However, sometimes a system cannot detect conflict during its run-time and later realizes that it happened because of delayed sensor information \cite{al2019iotc}.

None of the above research considers impact conflict detection of IoT services, and existing conflict detection approaches cannot be applied directly. Current frameworks recognize the relationship between actions (e.g., opening a window blind) and their environmental effects (e.g., changes in illumination) but fail to measure the actual impact. Moreover, understanding conflicts within these systems requires knowing the preferences of the people who use them, necessitating to estimate their preferences. These estimated preferences are then fed into the impact conflict detection module to assess a conflict situation. To the best of our knowledge, \textit{this work is the first attempt to quantify the impact, estimate preferences, and then develop a conflict detection approach that can capture impact conflicts with their associated likelihood values}.

\section{Conclusion and Future Work}
% \vspace{-4mm}

We propose a novel approach for impact conflict detection of IoT services by quantifying the impact and estimating preferences. The proposed impact assessment model is developed based on the integral of signal deviations strategy adopted from STL. We employ DBSCAN clustering algorithm to extract the residents' service usage preferences. Conflict likelihood is computed using preferential proximity and temporal proximity strategies. Experimental results show the effectiveness of the proposed approach.

While offering convenience, smart home IoT devices collect extensive personal data, raising substantial privacy and security concerns. The risk of data breaches and unauthorized access, such as hacking or malware, can compromise resident privacy and even allow physical intrusion into homes. Additionally, sharing collected data with third parties without consent highlights the privacy and ethical issues \cite{huang2023survey}. Despite the importance of robust security measures to protect user data, this research does not cover privacy and security considerations. Research topics, such as ``occupancy estimation'', ``resident identification'' play a crucial role in the smart home research domain \cite{qin2021occupancy, song2023positional}. However, addressing the challenge of attributing service events to specific users in a multi-user environment is beyond the scope of this study.

Future enhancements to our framework will involve integrating additional contextual data, such as interpersonal relationships, to improve conflict detection and resolution in smart homes. We aim to test our solutions in more complex scenarios using broader experimental setups and larger datasets. Moreover, we will examine these mechanisms' ethical and legal considerations to prioritize privacy, autonomy, and well-being. 

\def\IEEEbibitemsep{0pt plus 0.1pt}
\bibliographystyle{IEEEtran}
\bibliography{ImpactConflict}

\end{document}